# Signal Processing in the PVLAS Experiment

E. ZAVATTINI[1], G. ZAVATTINI[4], G. RUOSO[2], E. POLACCO[3], E. MILOTTI[-1],
M. KARUZA[1], U. GASTALDI[2], G. DI DOMENICO[4], F. DELLA VALLE[1],
R. CIMINO[5], S. CARUSOTTO[3], G. CANTATORE[1], M. BREGANT[1]
1 - Università e I.N.F.N. Trieste, Via Valerio 2, 34127 Trieste, ITALY
2 - Lab. Naz.di Legnaro dell'I.N.F.N., Viale dell'Università 2, 35020 Legnaro, ITALY
3 - Università e I.N.F.N. Pisa, Via F. Buonarroti 2, 56100 Pisa, ITALY
4 - Università e I.N.F.N. Ferrara, Via del Paradiso 12, 44100 Ferrara, ITALY
5 - Lab. Naz.di Frascati dell'I.N.F.N., Via E. Fermi 40, 00044 Frascati, ITALY
milotti@ts.infn.it   http://www.ts.infn.it/experiments/pvlas/

*Abstract:* - Nonlinear interactions of light with light are well known in quantum electronics, and it is quite common to generate harmonic or subharmonic beams from a primary laser with photonic crystals. One suprising result of quantum electrodynamics is that because of the quantum fluctuations of charged fields, the same can happen in vacuum. The virtual charged particle pairs can be polarized by an external field and vacuum can thus become birefringent: the PVLAS experiment was originally meant to explore this strange quantum regime with optical methods. Since its inception PVLAS has found a new, additional goal: in fact vacuum can become a dichroic medium if we assume that it is filled with light neutral particles that couple to two photons, and thus PVLAS can search for exotic particles as well. PVLAS implements a complex signal processing scheme: here we describe the double data acquisition chain and the data analysis methods used to process the experimental data.

## 1 Introduction

Because of the zero point fluctuations of charged particle fields (like the electronic field), the vacuum itself is expected to behave as material medium with dielectric properties. When an external magnetic field is present, quantum vacuum should behave as an optically anisotropic medium, and a linearly polarized field that propagates in a direction perpendicular to the field should undergo a (tiny) polarization change [1,2,3]

The PVLAS apparatus has been discussed elsewhere [5], and here we provide just a very short description: fig. 1 shows a horizontal schematic layout of the PVLAS ellipsometer: the actual set-up extends vertically. The interaction region, where infrared ($\lambda$ = 1064 nm) photons from a Nd-YAG laser interact with the external magnetic field, is contained within a high-finesse Fabry-Perot optical resonator consisting of a pair of dielectric, multilayer, high reflectivity mirrors ($M_1$ and $M_2$) placed 6.4 m apart. Vacuum (at a pressure less than $10^{-7}$ mbar) is maintained within a 4.6 m long, 25 mm diameter quartz tube (the tube and the additional vacuum chambers containing the optical elements are omitted in fig. 1 for clarity) traversing the room temperature bore of a 1.1 m long superconducting dipole magnet. A turntable, actuated by a hydraulic motor, rotates the whole liquid–He cryostat which contains the magnet [6], around a vertical axis, and thus provides a time-varying signal which is then used for heterodyne detection. During data taking, the turntable rotates at frequencies $f_M$ = 0.3–0.5 Hz. The heterodyne ellipsometer consists of a pair of crossed polarizing prisms P and A, together with an ellipticity modulator (Stress Optic Modulator, or SOM, see [7]).

The ellipticity modulator provides an ellipticity carrier signal for the purposes of heterodyne detection and is driven at a frequency $f_{SOM}$ = 506 Hz. A quarter–wave plate QWP can be inserted between FP cavity and SOM: when properly aligned it transforms apparent rotations (dichroisms) into ellipticities [8], which can then beat with the SOM carrier ellipticity signal and be detected. The laser beam, coming from a 1064 nm, 100 mW CW output power Nd:YAG laser, is kept at resonance with the FP cavity by means of an electro-optical feedback loop. This has the effect of amplifying the optical path within the interaction region by a typical factor of N = 63700. Details on the PVLAS cavity and on the frequency locking system can be found in [9]. Light transmitted through the analyser is detected by a photodiode and the resulting voltage signal is both directly fed to a computer for digitization at an 8.2 kHz sampling rate ("fast" acquisition) and to a lock-in amplifier referenced at the 506 Hz SOM carrier frequency: the demodulated signal is frequency analysed on-line by a spectrum analyser, and digitized for off-line analysis ("slow" acquisition). The triggering and gate signals for the slow acquisition are obtained from a series of 32 marks placed around the circumference of the turntable: in this way, for every acquired data point the direction of the rotating magnetic field with respect to the fixed initial polarization direction is known, and absolute phases can be determined for all acquired signals. The detection photodiode signal is analysed off–line giving amplitude and phase of quantities of interest. In particular, ellipticity and dichroism signals generated by interactions with the magnetic field should appear, in a Fourier spectrum, at twice the magnet rotation frequency (since data acquisition is synchronous with magnet rotation, this is the convenient unit for measuring frequencies). Fig. 2 shows a panoramic view of the experimental area.

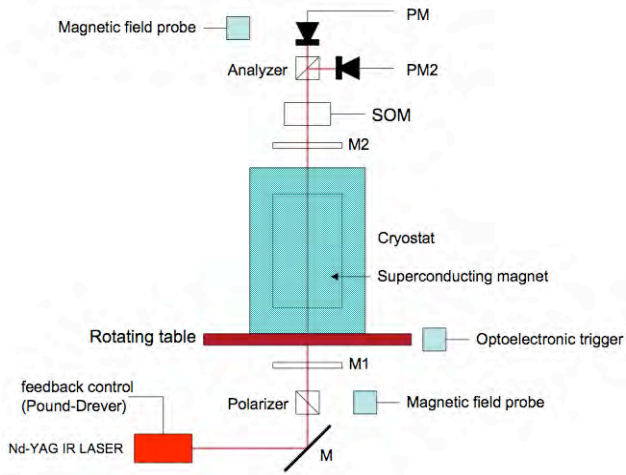

Fig. 1: Schematic layout of the PVLAS apparatus. The figure shows only the elements relevant to the discussion of the heterodyne ellipsometer (see text) and omits details on the frequency-locking feedback loop necessary to keep the laser and the FP cavity at resonance, the vacuum chambers, the cryogenic system, the magnet rotation assembly and the data acquisition and digitization system. PM = primary photodiode (extraordinary beam of Glan polarization analyzer); PM2 = normalization photodiode (ordinary beam); SOM = Stress Optic Modulator; M1, M2 = Fabry-Perot cavity mirrors. The figure also stresses the presence of some additional sensors (magnetic field probes, and most of all the optoelectronic trigger that provides the basic signal for both DAQ systems, both "slow" and "fast").

## 2  The PVLAS Signal

In an ellipsometer such as PVLAS it is easy to see that the photodiode signal is proportional to the square of the sum of the ellipticities associated to different optical elements:

$$I \approx I_0 \left( \sigma^2 + \left( \Psi(t) + \eta(t) + \Gamma(t) \right)^2 \right) \qquad (1)$$

where $I$ is the photodiode current, $\sigma^2$ is the extinction of the polarizer-analyzer pair, $\eta(t)$ is the ellipticity of the SOM, $\Gamma(t)$ is the residual and slowly drifting uncompensated ellipticity associated to the ellipsometer optics and $\Psi(t)$ is the physical signal that we wish to detect; $\eta(t)$ is modulated at 506 Hz, $\eta(t) = \eta_0 \cos(\varphi_{SOM}(t) + \varphi_{SOM,0})$, while $\Psi(t)$ is modulated at twice the magnet rotation frequency, $\Psi(t) = \Psi_0 \cos\left(2\varphi_M(t) + 2\varphi_{M,0}\right)$, where $\varphi_{SOM}(0) = \varphi_M(0) = 0$. When we expand equation (1), we find that the photodiode signal is a sum of several sinusoidal functions with the time-dependent arguments and constant phases listed in Table 1 [10].

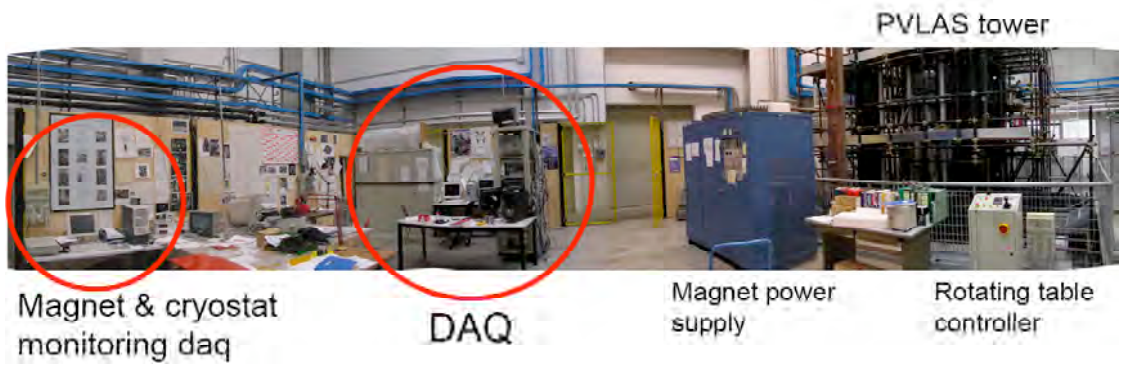

Fig. 2: panoramic view of the experimental area (november 2004). On the right the PVLAS tower is partly visible (the upper optics with the main photodiode is higher up, outside the visible area). The power supply feeds the magnet with a current greater 2000A, to reach fields as high as 6T. The DAQ hardware is assembled together, near the center of the picture.

## 3  Slow DAQ Analysis

Fig. 3 shows schematically the data acquisition layout for the "slow" DAQ. Since each data acquisition run starts at a given angular position of the rotating table, the constant phase in the table can be retrieved from the sampled photodiode signal. Synchronization of sampling with table rotation (the number of samples correspond to an integer number of table rotations) makes signal windowing unnecessary as there is no spectral leakage. This kind of sampling requires additional filtering to avoid aliasing, and frequency-dependent corrections must be carried out offline: since filtering is affected by table rotation frequency, these corrections introduce additional phase and amplitude jitter when the table rotation frequency is not constant. Moreover a straightforward FFT analysis of the sampled data contains artifacts because it cannot account for the slight deviations of tick spacing from perfect uniformity. These artifacts can, at least in principle, be avoided by Lomb-Scargle analysis or by similar fitting methods that take into account the uneven tick spacing [12], but there is no way to prevent phase jitter due to the coupling between the variable rotation frequency and the low-pass filter windows.

Table 1: main Fourier components of the PVLAS signal

| time dependent argument | constant phase | amplitude |
|---|---|---|
| $\approx 0$ | 0 | $\sigma^2 + \frac{1}{2}\Psi_0^2 + \frac{1}{2}\eta_0^2 + \Gamma^2(t)$ |
| $2\varphi_M(t)$ | $2\varphi_{0,M}$ | $2\Gamma(t)\Psi_0$ |
| $4\varphi_M(t)$ | $4\varphi_{0,M}$ | $\frac{1}{2}\Psi_0^2$ |
| $\varphi_{SOM}(t) - 2\varphi_M(t)$ | $\varphi_- = \varphi_{0,SOM} - 2\varphi_{0,M}$ | $I_- = \eta_0 \Psi_0$ |
| $\varphi_{SOM}(t)$ | $\varphi_1 = \varphi_{0,SOM}$ | $I_1 = 2\eta_0 \Gamma(t)$ |
| $\varphi_{SOM}(t) + 2\varphi_M(t)$ | $\varphi_+ = \varphi_{0,SOM} + 2\varphi_{0,M}$ | $I_+ = \eta_0 \Psi_0$ |
| $2\varphi_{SOM}(t)$ | $\varphi_2 = 2\varphi_{0,SOM}$ | $I_2 = \frac{1}{2}\eta_0^2$ |

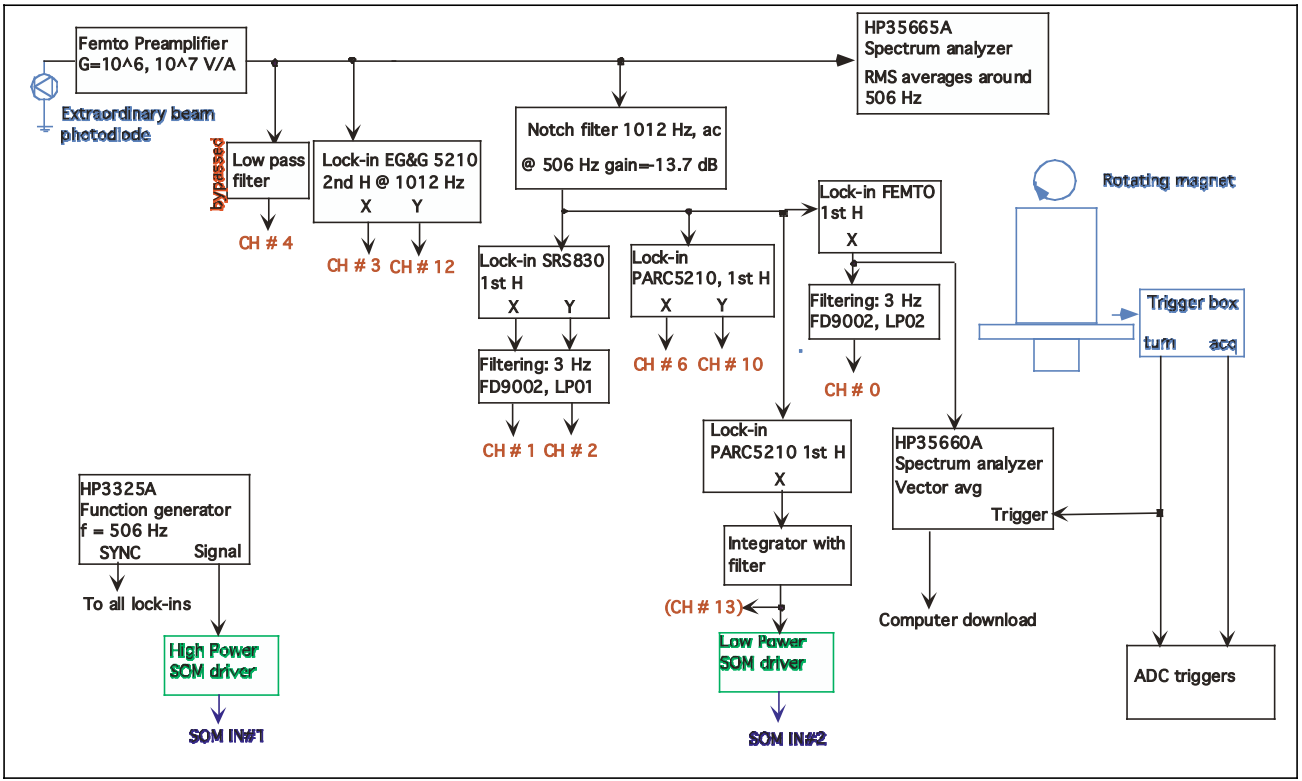

Fig.3: schematic layout of the "slow" DAQ. The light signal is converted to a current signal by the photodiode (upper left) and passed to a FEMTO current-to-voltage preamplifier. The voltage signal is analyzed by a spectrum analyzer, and passed to a couple of lock-in amplifiers. The clocking signal for the lock-in amplifiers is provided by the same oscillator that drives the SOM (lower left): the second harmonic lock-in provides the SOM amplitude $\eta_0$, while the first harmonic lock-in follows a notch filter at 1012 Hz, which reduces common mode effects. The output of the first harmonic lock-in is further low-pass filtered to prevent aliasing and digitized by an ADC triggered by logic signals from the rotating magnet (the DAQ trigger logic is omitted from this scheme for simplicity). Several other signals are acquired for monitoring purposes.

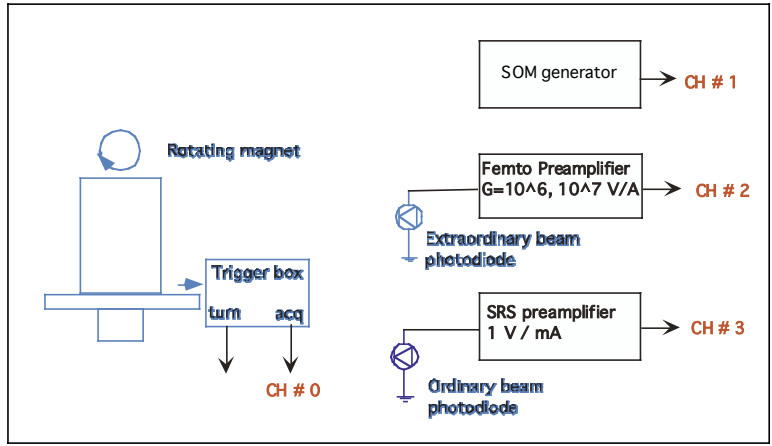

Fig 4: schematic layout of the "fast" DAQ. Four basic signals are sampled by a set of Sigma-Delta ADC's: the angular position signal provided by the positioning ticks on the rotating table, the output of the SOM driving oscillator, and the signals from the two main photodiodes.

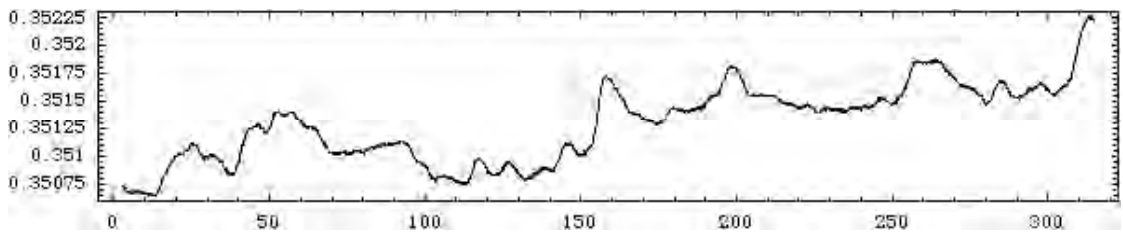

Fig. 5: table rotation frequency during PVLAS run 000807_0 (rotation frequency(Hz) vs. time (s)). These data have been obtained from the analysis of the tick signals digitized by the "fast" DAQ. This rotation frequency curve is numerically integrated to calculate the angular position of the rotating table and thus provides the $2\varphi_M(t)$ time dependent argument listed in the table.

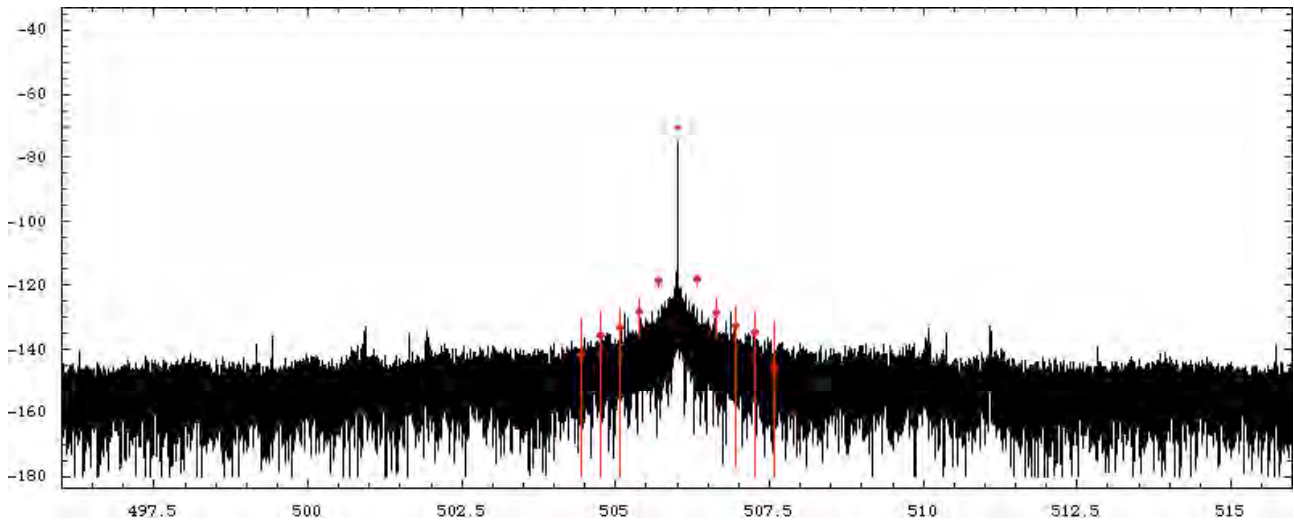

Fig. 6: FFT spectrum and fit results in PVLAS run 000892_0_1. The spectrum has been obtained from a windowed FFT (even though ticks are not evenly spaced the FFT is useful for visualization) on a total of $2^{24}$ samples, so that the frequency resolution is about 0.5 mHz over the whole range from 0 Hz to the Nyquist frequency (4.2 KHz). The dots show the results of the fitting procedure described in [13], which corrects for the instabilities of table rotation and for uneven tick spacing and tries to detect sidebands at integer multiples of the table rotation frequency (the number of samples in the fitting procedure is not necessarily a power of 2 and in this case there are 18989249 valid samples).

## 4 Fast DAQ Analysis

Fig. 4 shows the data acquisition layout for the "fast" DAQ: the hardware path for the "fast" DAQ is simpler, and all the complexity of the analysis is pushed into the software. Data are demodulated offline by specialized software that uses the tick signal to avoid all leakage [13]. Table rotation frequency is automatically accounted for and does not introduce additional jitter. Data files are sometimes very large (the largest datafiles analyzed to date take 2GB of disk space, corresponding to more than 67 million 8-channel samples), and the analysis may be quite slow (the largest datafiles – 2.5 hours of data taking time - take about 5 hours on a 1.6 GHz double processor Apple Macintosh G5). Fig. 5 shows the plot of the table rotation frequency during a single data acquisition run: this function has been reconstructed from the sampled tick signal. Fig. 6 shows a narrow frequency region around the SOM frequency: it is in this region that we expect to find the sidebands (see table) that carry the important physical information.

## 4 Conclusion

The PVLAS ellipsometer has been designed to measure the optical properties of quantum vacuum. Both data taking and data analysis are very challenging because of the smallness of quantum effects. Here we have provided a brief description of the data acquisition hardware and of the software programs that have been developed by the PVLAS collaboration. The experimental results shall be reported elsewhere [14].